\documentstyle[twocolumn, prl, aps, epsfig]{revtex}
\def\bea{\begin{eqnarray}}
\def\eea{\end{eqnarray}}
\begin{document}

\draft
\title{Corrections to scaling for percolative conduction: anomalous behavior at small L}
\author{Ivica Re$\check{s}$ \cite{res}}
\address{Department of Physics and Astronomy,
University of Kentucky,  Lexington, KY 40506-0055}
\maketitle

\begin{abstract}
 
Recently Grassberger has shown that the correction to scaling
for the conductance of a bond percolation network on 
a square lattice is a nonmonotonic
function of the linear lattice dimension 
with a minimum at $L = 10$, while this anomalous behavior is not present
in the site percolation networks.  
We perform a high precision numerical study of the bond 
percolation random resistor networks on the square, triangular
and honeycomb lattices to further examine this result.
We use the arithmetic, geometric and harmonic means to obtain
the conductance and find that the qualitative behavior does not change:
it is not related to the shape of the
conductance  distribution for small
 system sizes. 
We show that the anomaly at small $L$ is absent on the 
triangular and honeycomb networks.
We suggest that the nonmonotonic behavior is an artifact 
of approximating the continuous system for which the theory is
formulated by a discrete one which can be simulated on a computer.
We show that by slightly changing the definition of the linear 
lattice size we can eliminate the minimum at small $L$
 without significantly affecting the large $L$ limit.  

\end{abstract}


\section{INTRODUCTION}
According to finite size scaling theory,
the conducance of a finite random resistor network
is  expected to vary with the system size:
\bea
<\sigma>=L^{-t}[a + b f(L)]
\label{Eq:eq1}
\eea  
Here $t \approx  0.982$  \cite{gras} is the conductivity exponent 
and $f(L)$ is the correction to  
scaling term which vanishes as the size of the system
becomes infinite \cite{stauf,frank,lob}.In this work we study
the behavior of the correction-to-scaling for finite systems
that can be simulated on a computer.

Recently Grassberger \cite{gras} numerically studied the conductance
of random resistor networks on the  square lattice by numerical 
simulations for $2 \le L \le 4096$.
He found that for bond percolation networks  
 the corrections-to-scaling are nonmonotonic: 
there is a  dip in the corrections-to-scaling present at small lattice sizes.
Even though  this effect is rather weak, it can be clearly distinguished 
by high precision numerical simulations. 
In the site percolation networks this behavior is absent. 
Grassberger showed that the bond percolation  data could be fitted to the log-periodic form
 $<\sigma> L^{t} \sim \sin(\log(L))$ and pointed out that only the comparison 
with site percolation resistor networks enables us to reject this. 
Because this behavior is not understood  we wanted to 
investigate it further by studying systems defined on different
lattices.
One question we ask is whether this behavior is the result of the 
strong shape dependence of the probability distribution of conductances 
for small system sizes. To answer this we simulated square lattice
percolating networks  and calculated the arithmetic, 
geometric and harmonic means (Grassberger \cite{gras} used the arithmetic 
mean to obtain the conductance). Another question is whether the 
oscillations are universal or particular  to the square lattice used 
in \cite{gras}. To test the universality we also simulated bond 
percolation resistor networks on triangular and honeycomb lattices. 
  
Our paper is organized as follows: in Section 2 we give the details of our 
numerical work and present our results, Section 3 contains the  discussion 
and  conclusions.
 
\section{Numerical simulations }

The systems we study are random resistor networks that are connected to 
perfectly conducting busbars at the two opposite edges of the array, while
the transverse boundaries are free. 
To simulate a random resistor network we start with a bond percolation network 
and  assign unit conductance to the bonds that are present while a 
missing bond gives zero
conductance. The calculation of the network conductance for a given 
configuration of conductances is done by the Lobb-Frank algorithm \cite{lob},
which uses a succession of star-triangle (ST) and triangle-star (TS)
transformations to reduce the network to an equivalent chain of resistors.

Our square lattice simulations were performed on rectangles   
$(0 \le x \le L,1 \le y \le L)$,  with the perfectly conducting 
busbars placed at $x = 0$ and $x = L$. This means that the lattice 
(busbars excluded) has dimensions $(L-1,L)$. The reason for
this choice is to ensure that the bond system is self-dual (see Ref. 3
and references therein).
 To simulate systems on a triangular lattice we
add a diagonal resistor to each unit cell of the square lattice without
changing the position of the busbars. This way of representing the 
triangular network changes the geometry of the system - effectively 
the shape of the boundary becomes a rhombus \cite{ziff0}.
 Simulations on the honeycomb  lattice 
are done by starting with a honeycomb resistor network and replacing 
the leftmost column of resistors with perfect conductors and attaching the
rightmost 
column of nodes to a perfectly conducting busbar. The pictures of these networks 
can be found in Ref. 4.   
The systems were taken to be  at the percolation threshold, where 
$p_{c} = 0.5$ for square lattice,  $p_{c} = 2 \sin (\pi / 18) = 0.34729$ for the
triangular lattice and $p_{c} = 1 - 2 \sin (\pi / 18) = 0.65271$ for the
honeycomb lattice \cite{stauf,lajko}.
For the square lattice 
the conductance was obtained by exact enumeration for $L \le 4$, 
while for the triangular lattice exact enumeration was used to obtain 
the result for $L = 2$. For other system sizes the number of random 
configurations generated  was at least $10^{7}$.  

Our results for the square lattice are shown in Fig. 1, for the triangular lattice 
in Fig. 2 and for the honeycomb lattice in Fig. 3. 
We plot the conductance multiplied by $L$ raised to the conductivity 
exponent $t = 0.982$ as a function of $L$ on a semilogarithmic scale.
\begin{figure}

\centerline {\epsfxsize=3.in\epsfbox{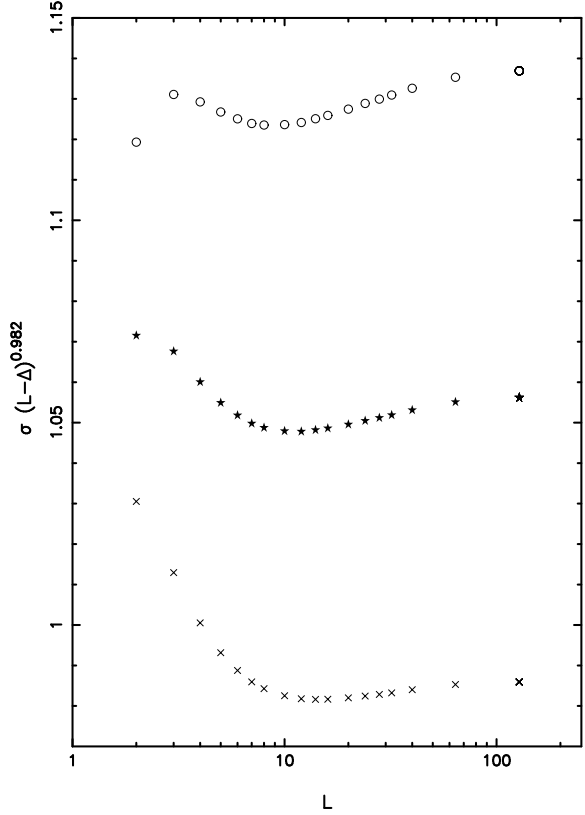}}
\vspace{3mm}
\caption{Average conductance on square lattices of size $L \times L$, multiplied by $L^{0.982}$.
  The upper set of data is obtained by using the arithmetic mean, the middle set is 
  obtained from the geometric mean and the lower set is the harmonic mean.
 }
\end{figure}
\begin{figure}
\centerline {\epsfxsize=3.in\epsfbox{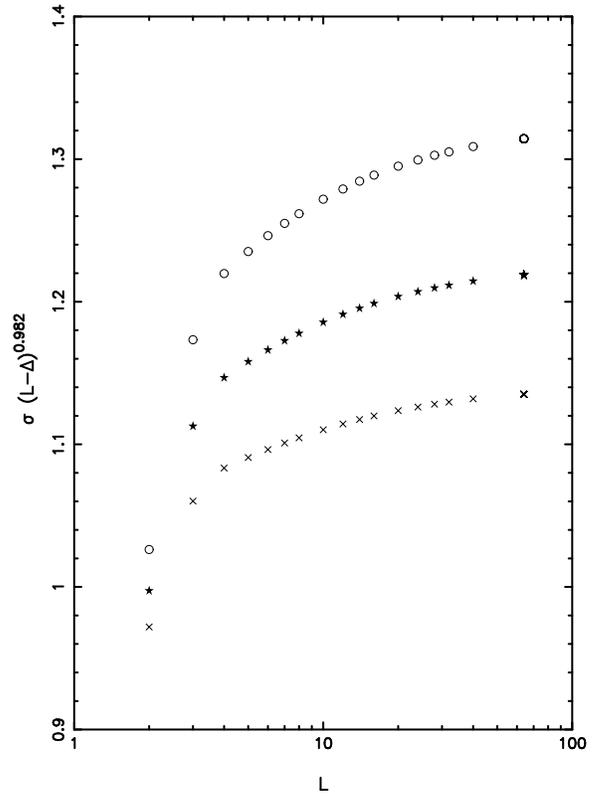}}
\vspace{3mm}
\caption{Average conductance on triangular lattices of size $L \times L$, multiplied 
    by $L^{0.982}$. The symbols have the same meaning as in Fig. 1}
\end{figure}
\begin{figure}
\centerline {\epsfxsize=3.in\epsfbox{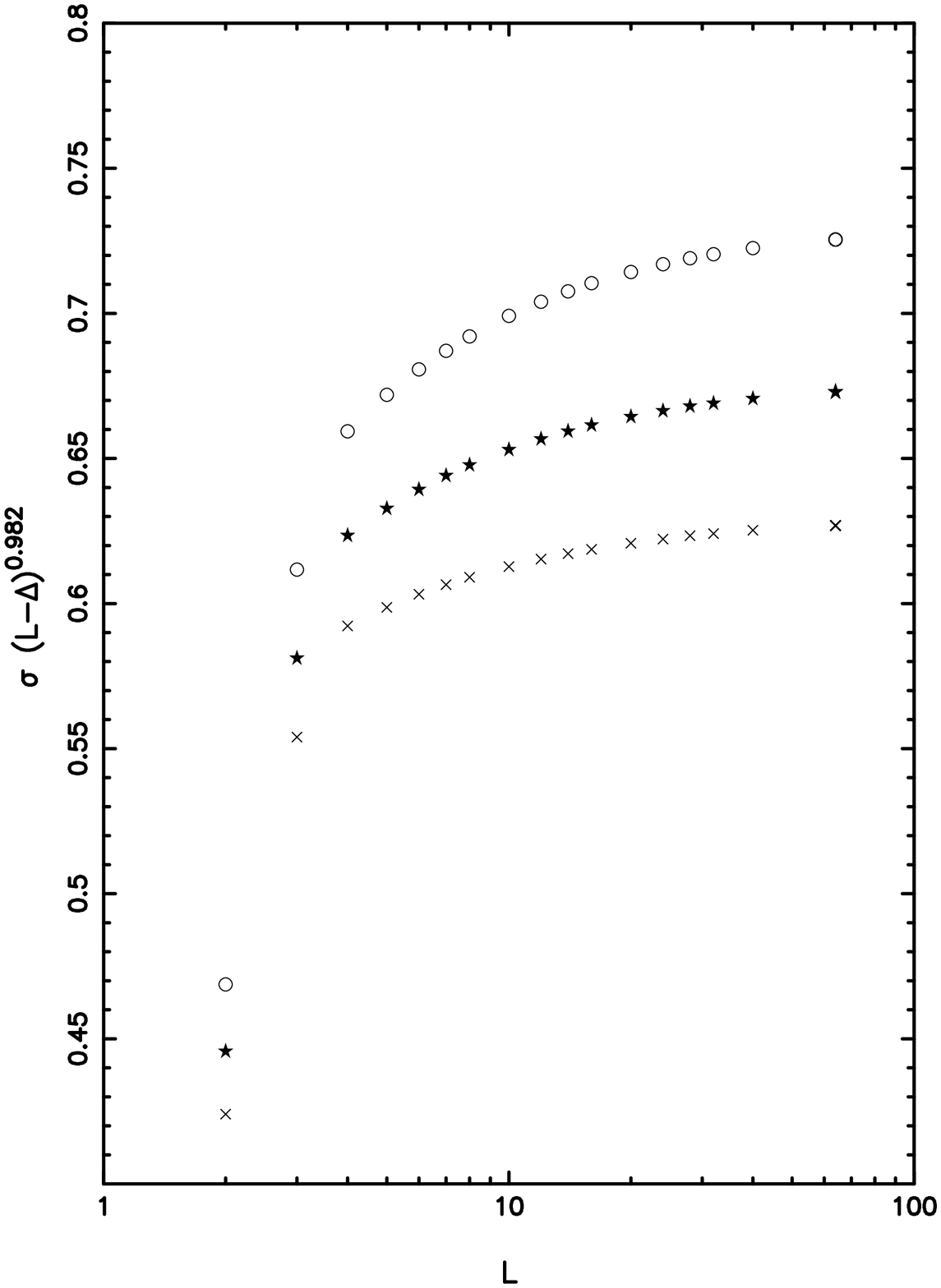}}
\vspace{3mm}
\caption{Average conductance on honeycomb lattices of size $L \times L$, multiplied 
    by $L^{0.982}$. The notation is the same as in Fig. 1 and Fig. 2.}
\end{figure}

In these figures full points (circles)  are obtained by using the arithmetic mean, 
stars are obtained from the  geometric mean and crosses correspond
to the harmonic mean (where the averages are computed using only the samples
with nonzero conductance). The errors are smaller than the size of the points
on the graphs. The points in Fig. 1 obtained by the arithmetic mean 
match the data obtained by Grassberger in \cite{gras}.
 
\section{Discussion and Conclusions}

Let us first consider our results for the networks defined on a square 
lattice.

 The data obtained by using the arithmetic mean start
at a low value for the lowest lattice size $(L=2)$ and then there is a jump 
to a high value at lattice size $L = 3$ followed by the monotonic decrease
up to $L\approx 10$.
If we use the harmonic or geometric mean the jump is not present -
 $<\sigma> L^{t}$ is a monotonically decreasing function until it reaches 
a minimum. Thus the behavior at $L=2$ is affected by the peculiar 
(and very discrete) distribution of conductances.

In all three cases a minimum  is present at low $L$.
Our numerical investigation of conduction on a square lattice 
shows that the dip in the correction-to-scaling term for small systems
does not get averaged out by using different means to obtain 
the conductance. 
 We did not find the anomalous behavior  for the systems defined on 
triangular and honeycomb lattices. The dip at small $L$ is not present 
and the behavior is quite similar to those obtained in Ref. 1
for the case of site percolation on a square lattice.

The anomaly present in the case of square 
lattice is rather weak - only high precision simulations are able to reveal 
it. We  notice that changing $L$ by a fraction of the lattice spacing will 
strongly affect the behavior of $<\sigma> L^{t}$ for small $L$ while the large
$L$ results will be practically unaffected. 
Replacing $L$ by $L-\Delta$ as the independent variable gives a 
monotonic function for $\Delta = 0.05$; for $\Delta = 0.1$
the corrections-to-scaling  behave qualitatively in the same 
way as  the results obtained for the other systems mentioned 
above (Ziff \cite{ziff} used a similar constant in connection with
crossing probability problems). This is illustrated in Fig. 4.  
  \begin{figure}
\centerline {\epsfxsize=3.in\epsfbox{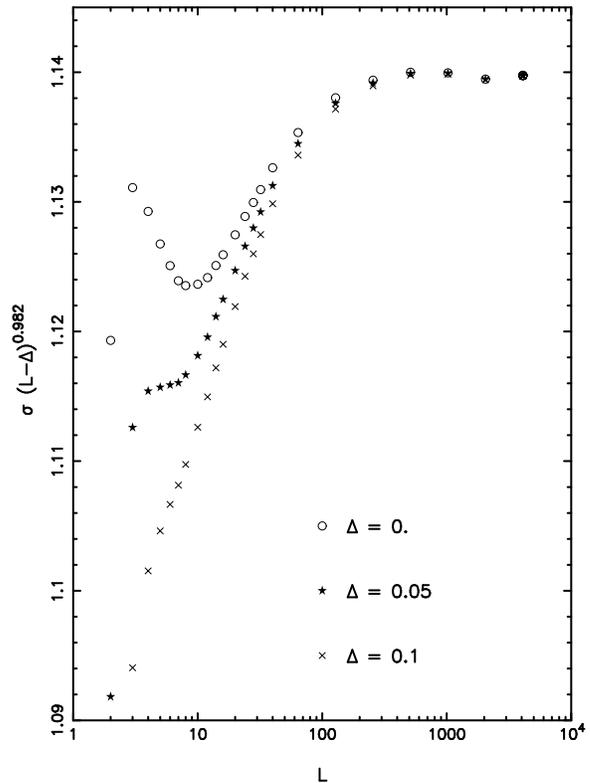}}
\vspace{3mm}
\caption{Average conductance on square lattices of size $L \times L$, multiplied by $(L-\Delta)^{0.982}$.
  The upper set of data is obtained by using the arithmetic mean (this set is represented in Fig. 1.
by the same symbol), the middle set is obtained by correcting $L$ by 0.05 and the lower set is 
obtained with a correction of 0.1. 
 }
\end{figure}
In the case of triangular and honeycomb systems the rescaling of $L$
does not qualitatively change the behavior of $<\sigma> L^{t}$ - in 
particular, the oscillation at small $L$ does not occur. This is shown 
in  Fig. 5 and Fig. 6 for triangular and honeycomb lattices, respectively.
  \begin{figure}
\centerline {\epsfxsize=3.in\epsfbox{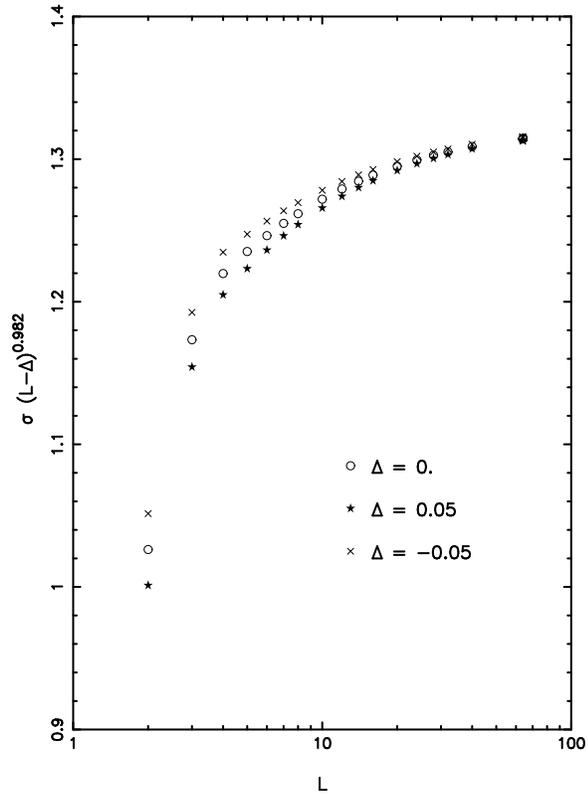}}
\vspace{3mm}
\caption{Average conductance on triangular lattices of size $L \times L$, multiplied by $(L-\Delta)^{0.982}$.
  The middle set of data is obtained by using the arithmetic mean (this set is represented in Fig. 2.
by the same symbol), the lower set is obtained by correcting $L$ by 0.05 and the upper set is 
obtained with a correction of -0.05. 
 }
\end{figure}
  \begin{figure}
\centerline {\epsfxsize=3.in\epsfbox{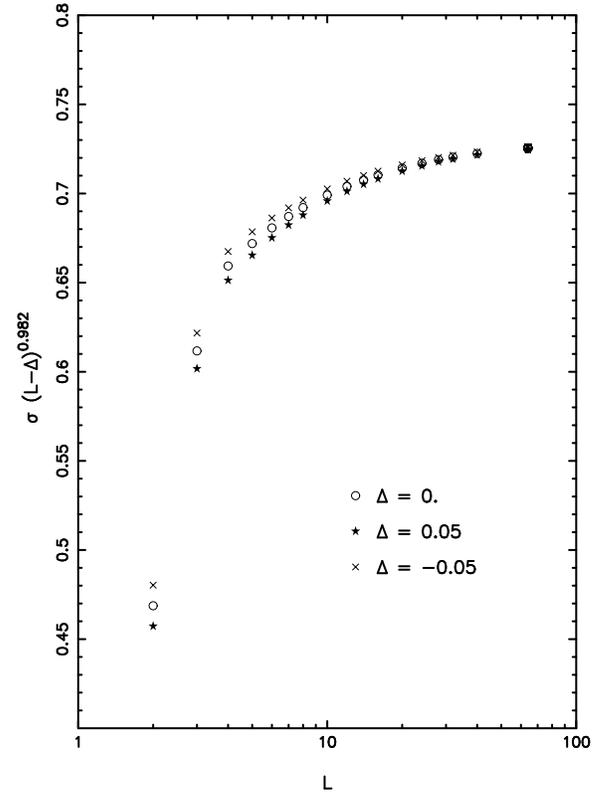}}
\vspace{3mm}
\caption{Average conductance on honeycomb lattices of size $L \times L$, multiplied by $(L-\Delta)^{0.982}$.
  The middle set of data is obtained by using the arithmetic mean (this set is represented in Fig. 3.
by the same symbol), the lower set is obtained by correcting $L$ by 0.05 and the upper set is 
obtained with a correction of -0.05. 
 }
\end{figure}
The justification for this adjustment of $L$ is that
the theory is formulated in the continuum limit while the discrete systems 
we simulated have the smallest scale equal to the lattice spacing. While 
for the large systems this should not matter, the small system results 
might be affected by this if we are doing very precise simulations.   

The author thanks J. P. Straley for critical reading of the manuscript.

\end{document}